Information Technologies in Public Administration

V. I. Gorelov

Russian Presidential Academy of National Economy and Public Administration

vigorelov@mail.ru

**Abstract**

There are visible changes in the world organization, environment and health of national conscience that create a background for discussion on possible redefinition of global, state and regional management goals. The author applies the sustainable development criteria to a hierarchical management scheme that is to lead the world community to non-contradictory growth. Concrete definitions are discussed in respect of decision-making process representing the state mostly. With the help of systems analysis it is highlighted how to understand who would carry the distinctive sign of world leadership in the nearest future.

In this work, I would like to offer the methodology, mechanism and road map for the actions which would enable implementation of programs for sustainable development of the humanity and various countries by means of information technologies.

First and foremost, application of information technologies in management requires creation of the model describing the condition of the administered item and the purpose of administration adequately. Administration itself is search for the optimum way of implementation of the set goal, control and forecasting. We are given an aggregate of tasks with ambiguous solutions. It is also relevant to development of the model, which is always approximation to reality, and selection of purpose and optimization.

Purpose.

The purpose of the research is sustainable development of the state. Sustainable development means sustainable development of the family, district, region, country and world – the principle of sub-region. It was obvious from the very beginning that territories of states are different in their area, mentality and geographical conditions, which means they must be administered differently, but the purpose at each level is the same.

Development of the Model.

In order to evaluate the state of the society, it needs to be structured first. In this case, structuring shall involve all the components: economic, social and environmental ones and it must be applicable to any size of society, starting from the family and up to the planet in general. Therefore, structuring must be unified. First of all, the units easily transformable for any level were determined by means of the typology. The table of conformity of family indicators and regional indicators is presented below as an example [3].

| Family condition indicators | Regional condition indicators |
|---|---|
| Total family income | Gross domestic product (GDP) |
| Total family expenditures | Budget |
| Family debt | Debt |
| Borrowing opportunities for development | Investment |
| Family poverty level | Poverty |
| Energy reserves | Internal energy |
| Raw stock | Extraction of raw materials |
| Need for energy reserves | Energy opportunities |
| Need for raw stock | Raw stock opportunities |
| Soil contamination level | Soil contamination level |
| Water contamination level | Scope of emissions into water |
| Air contamination level | Scope of emissions into the air (contamination) |
| Infrastructure of the family allotment | Infrastructure |
| Number of family members | Demography |
| Number of working family members | Working potential |
| Academic level | Education |
| Life quality | Life quality |
| Social morbidity | Social morbidity |
| Ecological morbidity | Pathologies |
| Family safety level | Not found |
| Level of use of education in the family | Scientists' remuneration |

It should be noted that transformation of indicators depends on the scale of the studied society. Therefore, the task of sustainable development administration is unified: uniform purpose for each level of the society (structure may change a little subject to the selected indicators). This approach allows creating the hierarchic system of consistent purposes and finding optimum strategic ways of development.

The next step is to select indicators characterizing the condition of each unit of the society. The point is that accurate analysis requires the model's indicators to draw upon quantitative rather than qualitative characteristics and, therefore, to be included into the official statistics [7-13].

The following indicators were selected from the WB data for 2000 [5]:

| Regional condition indicator | Selected indicator |
|---|---|
| GDP | GDP |
| Budget | Budget |
| Debt | Interest-bearing repaying to GDP ratio |
| Investment | Scope of investment (internal and external ones) into the Region |
| Poverty | Gini coefficient |
| Internal energy | Volume of produced energy in conventional tons |
| Extraction of raw materials | Volume of produced raw materials in monetary terms |
| Energy opportunities | Imported or exported energy to the used territory ratio |
| Raw stock opportunities | Ratio between raw stock export and import difference and the machine building and chemical industry production Scope |
| Soil contamination level | No statistical data found |
| Scope of water contamination | No statistical data found |
| Air contamination | Volume of emissions per square meter of the territory |
| Infrastructure | Length of highways per square meter of the territory |
| Demography | Population |
| Working potential | Number of working people |
| Education | Education level according to the UNESCO methodology |
| Life quality | Human development index (HDI) |
| Social morbidity | Number of tuberculous patients per 10,000 people |
| Pathologies | Number of babies born with chromosome or genetic changes per 1,000 babies – no statistical data found |
| Scientists' remuneration | Scientists' remuneration received for using solutions in their country |

The following indicators were selected from the Regions of Russia data for 2003 [6]:

| Regional condition indicator | Selected indicator |
|---|---|
| GDP | GRP |
| Budget | Budget |
| Debt | Regional budget balance |
| Investment | Scope of investment (internal and external ones) into the Region |
| Poverty | Gini coefficient |
| Internal energy | Volume of produced energy in conventional tons |
| Extraction of raw materials | Volume of produced raw materials in monetary terms |
| Energy opportunities | Imported or exported energy to the used territory ratio |
| Raw stock opportunities | Ratio between raw stock export and import difference and the machine building and chemical industry production scope |
| Soil contamination level | No statistical data found |
| Scope of water contamination | No statistical data found |
| Air contamination | Volume of emissions per square meter of the territory |
| Infrastructure | Length of railways and highways per square meter of the Territory |
| Demography | Population |
| Working potential | Number of working people |
| Education | Education level according to the UNESCO methodology |
| Life quality | Human development index (HDI) |
| Pathologies | Number of babies born with chromosome or genetic changes per 1,000 babies |

I would like to emphasize the extremely important indicator found in the data of the State Statistics Committee of Russia: the indicator of congenital abnormalities (birth defects) and chromosomal disorders with the first-listed diagnosis per 1,000 people. Environmental impact is slow and threshold. In other words, the body overcomes it for some time, but then the person falls ill. To my mind, the accumulated changes can be rather accurately characterized by their congenital transmission as these are established connections. This is why his indicator has been selected. In my opinion, this is the very indicator describing environmental impact on humans.

The selected indicators constitute the discrete cognitive system where each connection is defined by the coefficient of regression of impact of one standardized indicator on the other one. Each connection is checked in terms of the strong correlation dependence (at least 0.9), and the direction of connections is determined logically. If the correlation dependence is less than 0.9, the connection is excluded. Therefore, the rigorous cognitive model characterizing the condition of the society is developed.

Stages of system development can be briefly described as follows:

1. Creation of the database for each indicator.
2. Standardization of each column in the database and development of the standardized value matrix.
3. Consistent calculation of column correlation matrix, correlation error matrix, regression matrix, and regression error matrix.
4. Identification of correlation connections between columns with error therein not exceeding 0.1 (90 % reliability of the material connection).
5. Calculation of coefficients of material connection regressions on the basis of the regression matrix.
6. Development of the directed graph of the system describing mutual impact of one indicator on the other ones, by means of coefficients of material connection regressions (Table 1). In this case, the direction of connections is determined by logic (change of one indicator in the developed system shall affect any other one).

Table 1. Matrix of Adjacency of the Soft Cognitive System Model. The calculated regression coefficients were used in calculations instead of units.

| | **GDP** | **Education** | **Morbidity** | **Debt** | **Energy resources** | **Raw stock** | **Infrastructure** | **Investment** | **Energy opportunities** | **Raw stock opportunities** | **Demography** | **Poverty** | **Remuneration** | **Pathologies** | **Contamination** | **Budget** | **HDI** | **Workers** |
|---|---|---|---|---|---|---|---|---|---|---|---|---|---|---|---|---|---|---|
| **GDP** | | 1 | | | | | | | | | | | 1 | | 1 | 1 | 1 | |
| **Education** | | | 1 | | | | | | 1 | | | | | | | | 1 | 1 |
| **Morbidity** | | | | | | | | | | | | | | | | | 1 | |
| **Debt** | | | 1 | | | | | | | | | | | | | | | |
| **Energy resources** | 1 | | | | | 1 | | 1 | | | | | | | | | | 1 |
| **Raw stock** | 1 | | | | | | 1 | 1 | | | | | | | | | | 1 |
| **Infrastructure** | 1 | | 1 | | | | | 1 | | | | | | | 1 | | 1 | |
| **Investment** | 1 | | | | | | | | | | | | 1 | | | | | |
| **Energy opportunities** | | | | | | | | | | 1 | | | | | | | | |
| **Raw stock opportunities** | | | | | | | | | | | | 1 | | | | | | |
| **Demography** | | 1 | | | 1 | | | | | | | | | | | | | |
| **Poverty** | | 1 | 1 | 1 | | | | 1 | | | | | | | | | | |
| **Remuneration** | | 1 | | | | | | | | | | | | | | | | |
| **Pathologies** | | | | | | | | | | | 1 | | | | | | | |
| **Contamination** | | | | | | | | | | | | | | 1 | | | | |
| **Budget** | | 1 | 1 | | | | 1 | | | | | | | | | | | |
| **HDI** | | | | | | | | | | | | 1 | | | | | | |
| **Workers** | | | | | | | | 1 | | | 1 | | | | | 1 | | |

The next step is modeling of the process of changes. For this purpose, one signal is sent to each top of the model (indicator). It should be noted that the society is resistant to changes. Therefore, the adjacency matrix in the correctly developed system is the finite dimensional contraction operator. It means that the aggregate impact of the signal onto the system is finite and can be calculated.

As response to one signal from different tops of the model is different, they can be compared to each other. Estimation of the system weight of the signal can be found in works [1, 2, 3, 14, 15, 16]. It turned out that system weights allow determining the condition of the system (stagnancy, efficient management, pre-crisis condition etc.) [2] as well as the connection through which interaction is most efficient [2, 4] for this condition of the system and at the given moment on the basis of the modeling result. It allows controlling implementation, and consistent step-by-step modeling enables forecasting development. The forecasting horizon depends on the scope of the object. More detailed description with analysis of system weights is given in works [2, 3].

Table 2

Let's consider the calculated system weights for the overall world development for 2000 (calculations were made with the database of the World Bank for 2000 [5]).

| **Criterion** | **System weight** |
|---|---|
| Education | 7.27129 |
| Internal energy | 5.19433 |
| Extraction of raw materials | 3.77306 |
| Working potential | 3.11753 |
| Demography | 2.96224 |
| Life quality | 2.31044 |
| GDP | 2.23819 |
| Infrastructure | 1.56816 |
| Investment | 0.77448 |
| Raw stock opportunities | 0.70011 |
| Energy opportunities | 0.53863 |
| Country budget | 0.49471 |
| Scientists' remuneration | 0.12092 |
| Debt | -0.56875 |
| Morbidity | -0.6064 |
| Poverty | -2.96161 |

Table 3

System weights of impact of the criteria onto development of Russia
(calculations were made with the database of the Regions of Russia for 2003 [6]).

| System weight | Criterion |
|---|---|
| 4.543663981 | Working potential |
| 3.886394368 | Population |
| 3.82008217 | Energy and fuel resources |
| 2.345511434 | Education |
| 0.982541451 | Raw stock opportunities |
| 0.444773845 | Energy and fuel opportunities |
| 0.403407945 | Raw stock |
| 0.396657503 | Infrastructure |
| 0.251260442 | Life quality |
| 0.10248979 | GRP |
| 0.102312859 | Investment |
| -0.020949465 | Air contamination |
| -0.033494984 | Regional budget |
| -0.2267694 | Regional budget balance |
| -0.768342744 | Stratification of society |
| -0.929894654 | Pathological morbidity |

Here the operator characterizes positive or negative impact on sustainable development, and the system weight describes the nature of changes in development of the system in case of one-time impact on the respective indicator.

It should be noted that the modeling results for different sections of the database of the World Bank (developed and developing countries) have given quite a complete and interesting picture to analyze the global environment. The analysis results are presented in [2, 3].

Below are the main conclusions of analysis of the sustainable development model:

1. For each territory, regardless of its size, there is its own development priority system at each moment. Due to cultural, religious, traditional and geographic differences, each territory has its own specific ways of sustainable development.
2. Development of the model allows developing priority hierarchy for each territory and peaceful sustainable development of the world community [17].
3. “Economic miracle”, coincidence of priorities of the leadership elite and sustainable development priorities. The bigger the deviation from sustainable development priorities is, the smaller the projection of the elite's priorities on the sustainable development vector is, and the slower the growth is.
4. The task of public administration is maximum approximation of the elite's interests to the development priorities of any territory at each moment. If any state wants to survive and develop properly, it shall promptly support development priorities at each stage by means of legislative acts and close cooperation of all branches of power. The new administration style is necessary: to do things beneficial for the state and its population.

5. From this point of view, the hierarchic system of sustainable development is efficient democracy.
6. It should be noted that the political system of states is not taken into consideration in the model. The problem is not the political system itself. It is the extent to which this system implements the sustainable development vector. From my point of view, Iceland, Switzerland and China are best at it whereas the priority vector of the EU elite demonstrates considerable deviation from the sustainable development vector.
7. The report to the Club of Rome has recently been published [18]. The report is being discussed, but it seems it is not going to be used actively to create development projects of the world and various countries. However, I believe the real ecological situation in the world is more critical than it seems.

Let's try to demonstrate it.

The World Bank's data contain emissions into the atmosphere, but there is no information on environmental impact. Let's carry out the qualitative analysis of impact the environmental factor onto development of the global community. For this purpose, let's add to the system the regression dependence of impact of air contamination on birth of children with chromosome and genetic changes on the assumption that this dependence is not very different from the respective indicator in Russia. Then the system weight of pathological morbidity (ecological condition of the environment) in the world development can be estimated approximately.

Table 4.

Results of Approximate Estimates of System Weights of Influence of the Indicators on Development of the World Countries for 2000 with Account of the Environmental Component

| **Criterion** | **System weight** |
|---|---|
| Education | 5.773738 |
| Electric energy production | 3.842598 |
| Extraction of raw materials | 2.732703 |
| Working potential | 2.414056 |
| Demography | 2.241922 |
| Life quality | 1.859361 |
| Infrastructure | 1.242785 |
| GDP | 1.031482 |
| Raw stock opportunities | 0.710317 |
| Energy opportunities | 0.541175 |
| Investment | 0.511257 |
| Country budget | 0.446617 |
| Scientists' remuneration | 0.11129 |
| Morbidity | -0.51918 |
| Debt | -0.57984 |
| Contamination | -0.66376 |
| Pathologies | -1.44159 |
| Poverty | -2.28817 |

As you can see, the qualitative estimates demonstrate that the planet cannot bear the environmental pressure anymore, and gives back 44 % of contamination. Of course, this weight is considerably higher in the developing countries. I wish I were mistaken, but the planet already seems to be suffering from the environmental crisis. I would like to make accurate calculations, which is quite possible if there are data.

Conclusion.

The main question arising is what to do. I think the humanity is able to reduce the environmental pressure, but it takes immense joint efforts:

1. We need the new development ideology. The planet must be saved for future generations. The environmental crisis must be slowed down. In this regard, the latest report to the Club of Rome is very important [18]. If its ideology is accepted by all the countries, the process will be under way. There is a huge niche of application of the capital and human resources [1-4].
2. This ideology may be implemented as
   - There is a methodology for development and study of management models with account of the systematic approach [2].
   - The sustainable development model has been created as a first approximation [1, 3].
   - The consistent principal scheme for calculation of sustainable development of the world in general and its parts, for instance, countries or groups thereof, has been created and tested [3,4 ].
3. New quantitative statistical indicators characterizing the condition of different sustainable development units need to be introduced.
4. These indicators need to be included into the sustainable development model, i.e. further work on the model.
5. Sustainable development priorities can already be calculated (provided that there is a statistical database) for each country and the world in general (on the basis of the existing model) in order to elaborate recommendations for strategic sustainable development. Major efforts.
6. Development projects could be assessed to choose the best.
7. Implementation of these projects can be controlled.

All these things can be done, but it requires joint efforts of the people interested in life of future generations.

It is a difficult path of peaceful development. Will it be followed? I do not know. All the world development history is an endless row of wars. Especially now, during the oil and gas war of the elite of the USA and EU and very tense global political circumstances in the context of three crises: environmental, economic and social ones. Almost destroyed Livia, Syria and Ukraine, strong migration flows to the EU, humanitarian and environmental regional catastrophes, and pressure on the DPRK. It seems the world is approaching the global war. But there will be no winners in this war, losers only.

References.

[1] V.I. Gorelov. Cognitive model of sustainable development. 30th volume of "International Journal of Information Technology & Business Management", 2014. ISSN 2304-0777

[2] V.I. Gorelov, O.L. Karelova Системное моделирование в социально-экономической сфере. (System modeling in socio-economic sphere) Химки: РМАТ, 2012.-185с

[3] V.I. Gorelov. Управление развитием регионов. (Management of development of regions) Москва, Экон-Информ, 2007, 163с.

[4] P.E. Golosov, V.I. Gorelov. Notes on World’s Sustainable Development, 2015 ICCBES, ICCBES-742

[5] Статистический справочник «Страны и регионы. 2000» – М: изд-во Весь мир, 2001 -240 с.

(World Development Indicators -2000).

[6] Регионы России. Социально-экономические показатели, 2003. –М: изд-во Госкомстат, 2003 - 1001с.

[7] Anker R. Gender and Jobs: Sex Segregation of Occupations in the World. Geneva, International Labour Office. 1997.

[8] Booysen E. An Overview and Evaluation of Composite Indices of Development // Social Indicators Research. 2002, Vol. 59- p. 115–151.

[9] Bourguignon E., Chakravarty S. Multidimensional Measures of Poverty. Delta Working Paper. Paris, 1998, p. 98–112.

[10] Deleeck K., van den Bosch K., de Lathouwer L. Poverty and the Adequacy of Social Security in the EC: a Comparative Analysis; Aldershot: Averbury. 1992.

[11] Eedderke J., Klitgaard R. Economic Growth and Social Indicators: An Exploratory Analysis //Economic Development and Cultural Change. 1998, №46 (April)- p. 455–489.

[12] Eurostat. Analyzing Poverty in the European Community (a). -Eurostat News – Special Edition, 1990.

[13] Eurostat. Poverty in Figures. Europe in the Early 1980s' (b). -Luxembourg: EUROP, 1990.

[14] Evstegneev D.V., Ledashcheva T.N. About methods of alternatives ranging at the optimal development strategy chose for modelling system //V International congress on mathematical modelling. Book of abstracts, volume II.- Dubna, 2002. – p.147-148

[15] Evstegneev D.V., Ledashcheva T.N., Gorelov V.I. About complex system of territory evaluation //V International congress on mathematical modelling. Book of abstracts, volume II. - Dubna, 2002. – p. 146-147

[16] Gorelov V.I., Evstegneev D.V. Methodological bases of cognitive models construction in ecology //V International congress on mathematical modelling. - Dubna, 2002.- p. 78-86

[17] Golosov P.E., Gorelov V.I., Karelova O.L. Use of Information Technology in the Government of a State, Word Academy of science, engineering and technology conference proceeding (19), 2017

[18] Доклад Римского клуба. von Weizsaecker, E., Wijkman, A. Come On! Capitalism, Short-termism, Population and the Destruction of the Planet. — Springer, 2018. — 220 p.